# Hybridization of photon-plasmon modes in metal-coated microtubular cavities

Yin Yin[1,2,*], Shilong Li[1], Vivienne Engemaier[1], Silvia Giudicatti[1], Ehsan Saei Ghareh Naz[1], Libo Ma[1,*], Oliver G. Schmidt[1,2]

[1]*Institute for Integrative Nanosciences, IFW Dresden, Helmholtzstr. 20, 01069 Dresden, Germany*

[2]*Material Systems for Nanoelectronics, Technische Universität Chemnitz, Reichenhainer Str. 70, 09107 Chemnitz, Germany*

*Corresponding authors: Y.Y. (y.yin@ifw-dresden.de) and L.B.M. (l.ma@ifw-dresden.de)

**Abstract:**

The coupling of resonant light and surface plasmons in metal layer coated optical microcavities results in the formation of hybrid photon-plasmon modes. Here, we comprehensively investigate the hybridization mechanism of photon-plasmon modes based on opto-plasmonic microtubular cavities. By changing the cavity structure and the metal layer thickness, weakly, moderately and strongly hybridized resonant modes are demonstrated depending on the photon-plasmon coupling strength. An effective potential approach is applied to illustrate the hybridization of photon-plasmon modes relying on the competition between light confinement by the cavity wall and the potential barrier introduced by the metal layer. Our work reveals the basic physical mechanisms for the generation of hybrid modes in metal-coated whispering-gallery-mode microcavities, and is of importance for the study of enhanced light-matter interactions and potential sensing applications.

Optical whispering-gallery mode (WGM) microcavities have gained considerable interest due to their unique properties such as small mode volume, ultrahigh quality factor, low threshold lasing [1,2], ultrasensitive sensing [3-5], optical spin-orbit coupling [6], and compatibility with on-chip integration technologies [7-9]. The combination of dielectric microcavities and noble metal layers allows for the interaction of resonant light with surface plasmons, resulting in novel phenomena such as surface plasmon polariton WGMs and hybrid photon-plasmon modes [10-15]. In these opto-plasmonic microcavities, the metal layer coated on the microcavity is capable of confining intense electromagnetic fields of plasmons localized at the metallic surface, which can be exploited for enhanced light-matter interactions. In a hybridized photon-plasmon mode the energy is partially stored in the traditional optical field located inside of the dielectric cavity and in the plasmon-type field partly localized at the metal coating layer. In previous seemingly contradicting reports, the plasmon-type fields have been reported to occur either at the inner surface of the metal layer coated on a solid microtoroid cavity [10,11], or at the outer surface of the metal layer coated on a solid cylinder cavity and a hollow microtube cavity [12,13,15]. As a result, the exact mechanism of the plasmon-type field to occur at the inner or outer surface of a metal layer has remained unexplored, which is however of fundamental interest for the understanding of photon-plasmon hybridization and relevant to the design of opto-plasmonic cavities.

In the present work, the formation of hybridized photon-plasmon modes is comprehensively investigated in metal layer coated microtubular cavities. In this kind of opto-plasmonic cavity, the thicknesses of both the cavity wall and the metal layer provide the degree of freedom to tune the photon-plasmon coupling. Three types of photon-plasmon modes are identified as weakly, moderately and strongly hybridized modes showing plasmon-type field localized at the inner, both inner and outer and outer surface of the metal coating layer, respectively. An effective potential approach is used to illustrate the generation and transition of these kinds of hybrid modes based on the competition between light confinement in the cavity and the potential barrier induced by the metal layer. Our work reveals the basic physical mechanisms of hybrid modes which are also applicable to other types of metal-coated WGM microcavities.

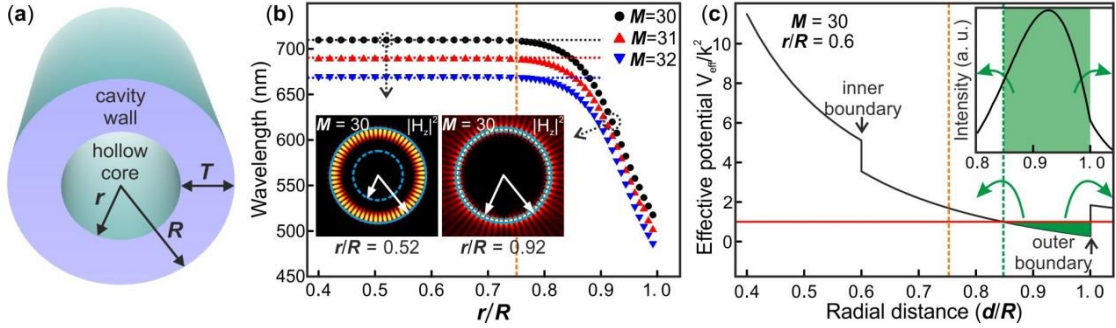

FIG. 1. (a) Sketch of a WGM microtubular cavity. *R*, *r* and *T* are the outer radius, inner radius and cavity wall thickness, respectively. (b) WGM resonant wavelengths (with azimuthal mode number *M* = 30, 31 and 32) as a function of *r*/*R*. Here, the outer radius *R* is fixed at 2.5 μm. The left and right inset indicate the corresponding field distribution in the cavity with *r*/*R* = 0.52 and 0.92, respectively. The yellow dashed line indicates the relative inner radius where the modes start to shift. (c) Effective potential distributed along the relative radial distance (*d*/*R*) for a representative microtubular cavity with *r*/*R* = 0.6 for *M* = 30. The inset shows the confined optical field distribution along the radial direction (*d*/*R*). The arrows denote the tunneling of confined photons to the classically forbidden region. The red line denotes the energy level of confined photons.

To investigate the confinement of the resonant light, a dielectric microtubular WGM cavity without any metal coating layer is introduced first. The microcavity consists of a hollow core and cavity wall, as sketched in Fig. 1(a). For simplicity, in the following study the outer diameter is fixed at *R* = 2.5 μm, and the refractive index is set as 1.6. The transverse magnetic (TM) modes (i.e. electric field perpendicular to cavity surface) are studied using the finite element method (FEM). In Fig. 1(b), the resonant wavelengths (with the azimuthal mode numbers *M*=30-32 as examples) are calculated as a function of *r*/*R*. In the range of *r*/*R* < 0.75, the resonant wavelengths remain constant, and the resonant modes are mainly confined by the outer boundary of the cavity. In this case, simple geometric optics or basic electromagnetic equations are sufficient to simulate the resonant wavelength neglecting the inner boundary, which is equivalent to the case of solid cylinder cavities. For propagating light reflected by the outer cavity boundary, the resonance occurs when $2\pi R\beta + \Phi = 2\pi[(M+0.5)+1.856(M+0.5)^{1/3}]$ [16], in which the propagation constant $\beta = n_{cavity}k_0$ ($n_{cavity}$ = 1.6), $\Phi$ is the extra phase caused by total internal reflections, **R** is the outer radius and **M** is the mode number. The calculated results are shown by dotted lines in Fig. 1(b), in agreement with the FEM calculations for the thick-walled cavities (**r**/**R** < 0.75). As **r**/**R** increases (> 0.75), the resonant wavelengths start to blueshift. In this regime, the conventional resonant condition mentioned above is no longer available because the light is effectively "reflected" by both outer and inner boundaries acquiring more complex phase corrections related to total internal reflections. Instead, the waveguide approach or the Mie scattering method are suitable for the calculation of optical resonances in thin-walled cavities [17,18].

The above analysis can be understood by employing an effective potential approach [19]. The radial optical field distribution in a microcavity can be described by a quasi-Schrödinger equation: $\nabla_r^2 \psi(r) + V_{eff}(r)\psi(r) = E\psi(r)$. This equation provides a classical interpretation of the electromagnetic wave behavior in analogy to the Schrödinger equation. In this approach the effective potential reads $V_{eff}(r) = k_0^2[1-\varepsilon(r)] + (\frac{M}{r})^2$ and the eigenenergy is $E = k_0^2$, where $k_0 = 2\pi/\lambda$ is the wave number in vacuum, $\varepsilon(r)$ is the permittivity along the radial direction and **M** is the azimuthal mode number. Figure 1(c) depicts the effective potential (for **M** = 30) in a microtubular cavity with **r**/**R** =0.6. The green zone in the potential indicates the regime where photons are classically allowed to be confined and have positive quasi-kinetic energy ($E > V_{eff}$). Obviously, the inner cavity boundary (**r**/**R** =0.6) can hardly influence the photons confined in the green zone when the cavity wall is thick enough. This potential approach well explains the constant resonant wavelengths in thick-walled cavities, where the resonant light is only confined by the outer cavity boundary and is insensitive to the inner boundary. One should note that the inner boundary of the green zone is located at **r**/**R** =0.85 (green dashed line) while the resonant wavelength starts to change when **r**/**R** reaches 0.75 (yellow dashed line, determined by rigorous solution shown in Fig. 1(b)). This phenomenon shows that the photons confined in the green zone start to "feel" the inner boundary at **r**/**R** =0.75 due to the tunneling effect as marked by the green arrows. Similarly, the evanescent field at the outer boundary can also be understood as a result of the tunneling effect through the outer boundary [19].

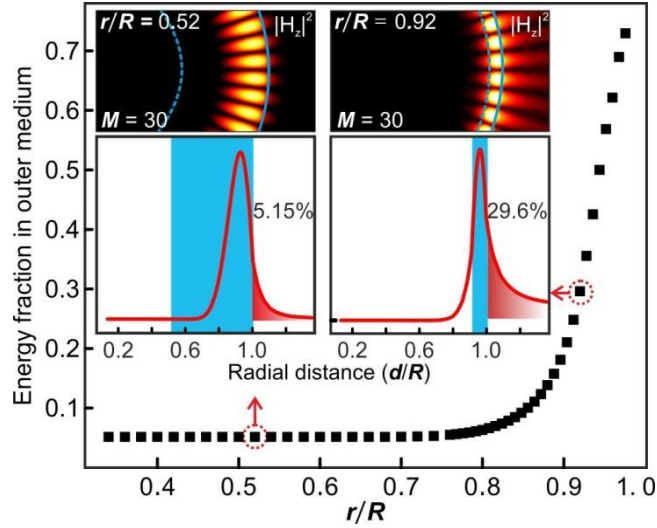

FIG. 2. Energy fraction in the outer surrounding of the cavity as a function of *r*/*R*. The left and right insets show the field distributions along the radial direction in the cases of *r*/*R* = 0.52 and 0.92, respectively.

The evanescent field of WGM cavities determines the strength of light-matter interactions and the sensitivity in detection applications. Especially the field strength plays a key role in the formation of hybrid photon-plasmon modes when interacting with the metal coating layer [15]. The evanescent field at the outer boundary of the dielectric microtubular cavity is characterized by the energy fraction found at the cavity surface, and is shown in Fig. 2 as a function of *r*/*R*. The energy fraction remains constant when *r*/*R* < 0.75 with a small value of 5.15%. As *r*/*R* > 0.75, the energy fraction exponentially increases when the cavity wall thickness decreases. For example, the energy fraction increases to 29.6% when *r*/*R* reaches 0.92, as shown in the right inset of Fig 2. The enhancement of the outer evanescent field as the cavity wall becomes thinner can be easily explained by the tunneling effect. For the cavity with the thick wall (i.e. smaller *r*/*R*), the resonant photons are well confined inside the cavity and have low kinetic energies. As a result, the tunneling probability is weak, and so is the evanescent field. In a thin-walled cavity, however, the photons can more easily tunnel out of the cavity boundary because of the higher kinetic energy, consequently leading to a stronger evanescent field. The difference in evanescent field intensities results in different interaction strengths, and thus different hybridizations when the light interacts with a metal coating layer – which will be discussed in the following.

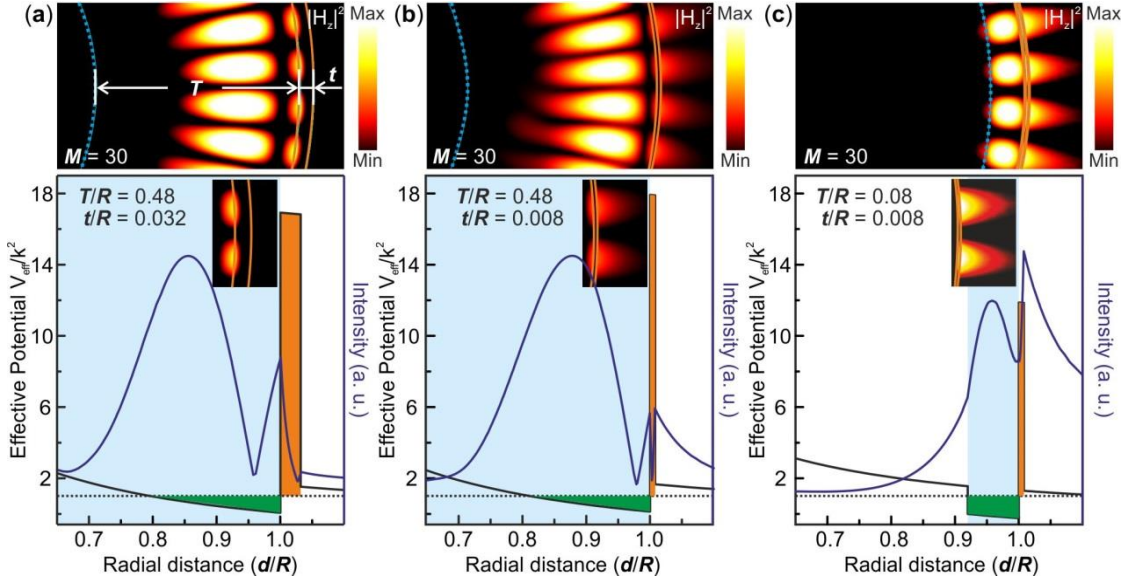

FIG. 3. Field distributions and effective potentials along the radial direction of (a) weakly, (b) moderately and (c) strongly hybridized photon-plasmon modes. The thicknesses of the cavity wall (*T*) and the metal coating layer (*t*) are set as *T*/*R* = 0.48 and *t*/*R* = 0.032, *T*/*R* = 0.48 and *t*/*R* = 0.008, *T*/*R* = 0.08 and *t*/*R* = 0.008, respectively.

The gold layer is used to coat the microtubular cavity for the generation of hybrid photon-plasmon modes, where the permittivities as function of optical wavelength are taken from Ref. [20]. By changing the degree of freedom for both the light confinement in the cavity wall and the gold layer thickness, three types of hybrid photon-plasmon modes are obtained. For instance, in a microtubular cavity with a wall thickness *T*/*R* = 0.48, the plasmon-type field of the hybridized mode is predominantly located at the inner surface of the metal layer when the gold layer thickness *t*/*R* = 0.032, as shown in Fig 3(a). As the main field energy is stored in the photonic mode part confined in the dielectric cavity, this mode is denoted a weakly hybridized photon-plasmon mode. When decreasing the gold layer thickness down to *t*/*R* = 0.008, an external plasmon-type field, in addition to the inner one, is formed at the outer surface of the gold layer, as shown in Fig. 3(b). Compared to the internal field shown in Fig. 3(a), the plasmon-type field located at the inner surface decreases and transfers to the outer surface. In this case, the plasmon-type field of the hybridized mode stays at both the inner and outer surfaces of the metal coating layer, and is regarded as a moderately hybridized photon-plasmon mode. Decreasing the cavity wall thickness (e.g. decrease to *T*/*R* = 0.08) while maintaining the gold layer thickness leads to a plasmon-type field of the hybridized mode which can be only seen at the outer surface of the metal layer, manifesting a strongly hybridized photon-plasmon mode as experimentally demonstrated in our recent work [15]. The absence of the inner plasmon-type field is explained by the fact that the strong evanescent field at the thin cavity wall surface directly tunnels out of the metal layer without interacting with the inner boundary.

The competition between light confinement in the cavity and potential barrier introduced by the metal layer determines the formation of these kinds of hybrid modes, which can be explained by the effective potential approach [11,15]. Potentials of three representative hybrid modes are shown in the bottom panels of Fig. 3(a)-(c). The green zones indicate the region where the resonant photons are confined. In the potential approach model, the deeper potential well confines photons with higher kinetic energy, which have a larger capability of tunneling out through the potential. It is obvious that the cavities in Fig. 3(a) and (b) have the same wall thickness, thus the two potential wells experience almost the same depth, indicating the same kinetic energy of the confined photons. In contrast, the thin-walled cavity in Fig. 3(c) has a deeper potential well confining photons with higher kinetic energy. On the other hand, the metal layer forms a potential barrier due to the negative permittivity. Both the width and height of the barrier influence the tunneling probabilities of the photons. The barrier width is determined by the thickness of metal layer, and the height is determined by the permittivity for the corresponding wavelength. However, for the first cavity in Fig. 3(a), the resonant photons confined in the potential well cannot tunnel out through the outer boundary as it is effectively screened by the high and wide metal barrier. Hence, only an internal field located at the inner metal surface is formed. In the second cavity in Fig. 3(b) the potential barrier is narrower, thus the photons can tunnel out into the outer surface, generating the extra external field located at the outer surface. For the third cavity in Fig. 3(c), due to the high kinetic energy of photons as well as the low metal barrier, the photons can directly tunnel out into the outside surface, resulting in the strongly hybridized external field.

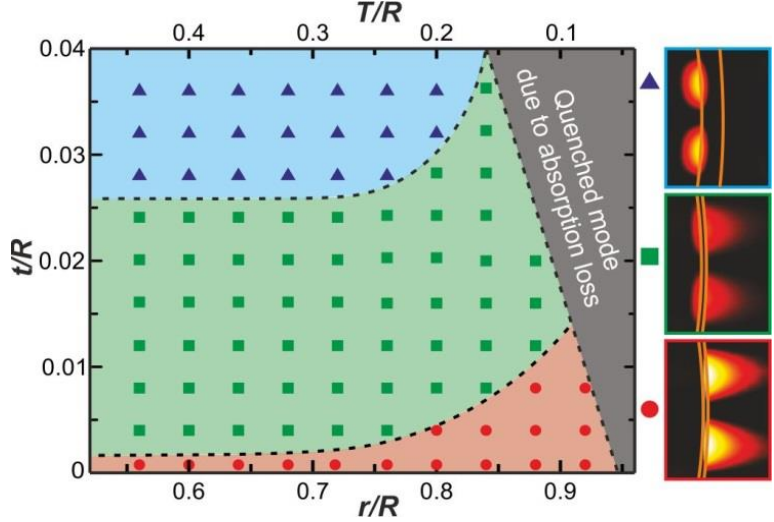

FIG. 4. Diagram of three types of hybrid modes in microtubular cavities with different thickness of cavity wall and metal coating. The triangles, squares and circles represent the weakly, moderately and strongly hybridized photon-plasmon modes, respectively. The typical field distributions of three hybrid modes are shown in the right insets.

The formation of the three types of hybrid modes is systematically investigated as a function of thickness of cavity wall and metal coating layer [see Supplemental material]. The hybrid modes are classified based on the field strength difference on the inner and outer surfaces of the metal coating layers. For the weakly hybridized mode, the internal field strength is more than an order of magnitude higher than the external field on the metal surface. When the external field intensity exceeds the internal one by an order of magnitude, the mode is defined as strongly hybridized. For the other situations, they are recognized as moderately hybridized modes, in which the plasmon-type fields can be found on both the inner and outer metal surfaces. All calculated three types of hybrid modes are respectively denoted by triangles, squares and circles, as shown in Fig. 4. The three zones are separated by dashed black curves. The two black curves become flat for the region $r/R < 0.75$, which is in agreement with the evolution behavior discussed in Fig. 1 and 2. On one hand, with increasing cavity wall thickness, light is better confined inside the dielectric cavity, making it hard to tunnel through the metal layer. On the other hand, the thickness of the metal coating layer also plays an important role in the generation of hybrid modes. As the coating layer becomes thicker, its shielding effect becomes more significant. In addition, the intrinsic absorption loss of the metal layer cannot be neglected, which reduces the Q-factor of the hybrid modes and renders the modes hardly detectable for thick metal layers, which is marked by the gray zone in Fig. 4.

In a recent experimental report [13], hybrid modes in a metal-coated microcylinder cavity have been discussed, where the field distribution was similar to that obtained in thick-wall microtubular cavities coated with thick metal layers (see Fig. 3(a)). This similarity is consistent with the above analysis where the inner boundary does neither affect the optical resonance nor the evanescent field at the outer boundary when the cavity wall is thick enough. However, if a hollow core is generated inside the microcylinder, e.g. by means of acid etching or vacuum assisted filtration [21-23], a transition to the other two types of hybrid modes with field distributions similar to those in Fig. 3(b) and (c), can be expected. Rolled-up nanotech is a very efficient way to create microtubular cavities with ultrathin walls by rolling up prestrained nanomembranes [17,24,25]. The cavity wall can be scaled down to 100 nm which supports WGMs with strong evanescent fields. And indeed, strongly hybridized photon-plasmon modes have been observed in the rolled-up microtubular cavities [15].

In conclusion, hybridizations of photon-plasmon modes in metal-coated microtubular cavities have been investigated. By changing the degree of freedom for both the thicknesses of the cavity wall and the metal coating layer, the strength of photon-plasmon coupling is tuned, giving rise to weakly, moderately and strongly hybridized photon-plasmon modes. The formation of such hybrid modes is illustrated by an effective potential approach, where the competition between light confinement in the potential well and potential barrier induced by the metal layer determines the field distribution of the hybrid modes. Our work provides a universal picture for understanding the basic physical mechanisms of photon-plasmon mode hybridization in metal-coated WGM microcavities, and is relevant for opto-plasmonic cavity designs. As a novel type of opto-plasmonic microcavity, the metal-coated microtubular cavity is promising for both fundamental and application-oriented studies such as enhanced light-matter interactions.